\documentclass[
 superscriptaddress,
reprint,
 bibnotes,
 showkeys,
 amsmath,amssymb,
 aps,
prb,
]{revtex4-1}
\usepackage{graphicx}
\usepackage{amsmath}
\usepackage{dcolumn}
\usepackage{bm}
\usepackage{braket} 
\usepackage{bbold}
\usepackage{diagbox}
\usepackage{multirow}
\usepackage{url}
\usepackage{color}
\usepackage[
 colorlinks=true,
 urlcolor=blue,
 citecolor=blue,
 linkcolor=blue,
 bookmarks=false,
 pdfstartview={FitH},
]{hyperref}

\begin{document}

\title{Raman scattering study of spin-density-wave-induced anisotropic electronic properties in $A$Fe$_2$As$_2$ ($A$ = Ca, Eu)}
\author{W.-L. Zhang}
\email{wlzhang@iphy.ac.cn}
\affiliation{Beijing National Laboratory for Condensed Matter Physics and Institute of Physics, Chinese Academy of Sciences, Beijing, 100190, China} 
\affiliation{Department of Physics $\&$ Astronomy, Rutgers University, Piscataway, New Jersey 08854, USA} 
\author{Z. P. Yin}
\affiliation{Department of Physics $\&$ Astronomy, Rutgers University, Piscataway, New Jersey 08854, USA}
\author{A. Ignatov}
\affiliation{Department of Physics $\&$ Astronomy, Rutgers University, Piscataway, New Jersey 08854, USA}
\author{Z. Bukowski}
\affiliation{Laboratory for Solid State Physics, ETH Zurich, 8093 Zurich, Switzerland}
\affiliation{Institute of Low Temperature and Structure Research, Polish Academy of Sciences, P.O. Box 1410, 50-422 Wroc\l{}aw, Poland}
\author{Janusz Karpinski}
\affiliation{Laboratory for Solid State Physics, ETH Zurich, 8093 Zurich, Switzerland} 
\author{Athena~S.~Sefat}
\affiliation{Materials Science and Technology Division, Oak Ridge National Laboratory, Oak Ridge, Tennessee 37831-6114, USA}  
\author{H.~Ding}
\affiliation{Beijing National Laboratory for Condensed Matter Physics and Institute of Physics, Chinese Academy of Sciences, Beijing, 100190, China} 
\affiliation{Collaborative Innovation Center of Quantum Matter, Beijing, China} 
\author{P. Richard}
\affiliation{Beijing National Laboratory for Condensed Matter Physics and Institute of Physics, Chinese Academy of Sciences, Beijing, 100190, China} 
\affiliation{Collaborative Innovation Center of Quantum Matter, Beijing, China}
\author{G. Blumberg}
\email{girsh@physics.rutgers.edu}
\affiliation{Department of Physics $\&$ Astronomy, Rutgers University, Piscataway, New Jersey 08854, USA}
\affiliation{National Institute of Chemical Physics and Biophysics, Akadeemia tee 23, 12618 Tallinn, Estonia}
 

\begin{abstract}
We present a polarization-resolved and temperature-dependent Raman scattering study of $A$Fe$_2$As$_2$ ($A$ = Ca, Eu). In the spin-density-wave phase, spectral weight redistribution is observed in the fully symmetric and non-symmetric scattering channels at different energies. 
An anisotropic Raman response is observed in the fully symmetric channel in spontaneously detwinned CaFe$_2$As$_2$ samples. We calculate the orbital-resolved electronic structures using a combination of density functional theory and dynamical mean field theory. We identify the electronic transitions corresponding to these two spectral features and find that the anisotropic Raman response originates from the lifted degeneracy of the $d_{xz/yz}$ orbitals in the broken-symmetry phase.

\end{abstract}

\pacs{74.25.nd, 74.20.Pq, 74.70.Xa, 75.30.Fv} 


\maketitle


The parent and underdoped compounds of the 122 family ($A$Fe$_2$As$_2$) of iron-pnictide superconductors harbor an antiferromagnetic ground state with a collinear stripe spin order of Fe ions. The formation of a spin-density-wave (SDW) order at a temperature $T_{SDW}$ is accompanied by a tetragonal to orthorhombic structural distortion in which the $C_4$ rotational symmetry is broken~\cite{Stewart_RMP2011}.
Although the lattice distortion is small, the electronic anisotropy can be large~\cite{Kasinathan_NJP2009}. For a strain-free sample, a twin domain structure with a typical dimension of a few micrometers usually forms with orthogonally aligned antiferromagnetic directions~\cite{Prozorov_PRB2009}. In many macroscopic symmetry-sensitive experiments, the response along the antiferromagnetic and ferromagnetic directions can be mixed and averaged. However, strongly anisotropic responses in the dc and ac electrical conductivity~\cite{Fisher_Science2010,Dusza_EPL2011,Uchida_PRL2012}, the thermoelectric power~\cite{Gegenwart_PRL2013} and the magnetic susceptibility~\cite{Matsuda_Nature2012} along the two orthogonal directions, as well as in the on-site energy splitting of the $d_{xz}$ and $d_{yz}$ orbitals~\cite{YiM_PNAS2011}, have been reported for samples completely or partially detwinned following the application of a uniaxial strain or stress~\cite{Tanatar_PRB2010,Fisher_Science2010}, or an in-plane magnetic field~\cite{Chu_PRB2010}.

Electronic Raman scattering is an inelastic light scattering process that traces the density-density correlation function driven by the incident and scattered lights. 
With a proper choice of light polarization, it can separate states or collective excitations associated with certain symmetries belonging to the irreducible representations for the point group of a particular crystal~\cite{Devereaux_RMP2007}. For example, the $d$-wave superconducting gap in cuprates results in a $\omega^3$ Raman response in the nodal direction and a linear Raman response in the anti-nodal direction~\cite{Devereaux_PRB1995,Tacon_nphys2006}.
In the iron-pnictides, the nematic fluctuations of the $XY$ symmetry are observed in the tetragonal phase with cross-polarized light along the Fe-As directions~\cite{Gallais_PRL2013,YXYang_JPS2014,Hackl_arXiv1507,Thorsmolle_arXiv_2014,WLZhang_arxiv2014}. 
Therefore, it is of interest to study the Raman response for excitations of different symmetries across the structural and magnetic phase transitions.

In this paper, we present polarization-resolved electronic Raman measurements of the iron-pnictide parent compound $A$Fe$_2$As$_2$ ($A=$ Ca, Eu). 
We observe two spectral features with different symmetries below $T_{SDW}$. One feature is a coherence peak in the fully symmetric channel that was previously assigned to the formation of a SDW gap near the $M(\pi/a,0,0)$ point~\cite{Chauviere_PRB2011}. Using symmetry analysis of a detwinned sample, as well as a combination of density functional theory and dynamical mean field theory (DFT+DMFT) calculations~\cite{Kotliar_RevModPhys, Haule_PRB2010} above and below $T_{SDW}$, we show that this feature is an intra-orbital transition near the $Z(0,0,2\pi/c)$ point. In the non-symmetric channel, we identify an additional small peak at lower frequency that originates from the lifted degeneracy of the $d_{xz}$ and $d_{yz}$ orbital at the $\Gamma$ point.  

The CaFe$_2$As$_2$ and EuFe$_2$As$_2$ samples employed in this study have been synthesized by Sn and Fe-As flux methods described in Refs.~\cite{Matusiak_PRB2010, Sefat_PRL2008}. CaFe$_2$As$_2$ goes through a second-order 
phase transition at 170 K whereas EuFe$_2$As$_2$ goes through a first-order 
phase transition at 175 K. The high-temperature tetragonal structure belongs to the space group I4/mmm (point group $D_{4h}$) and the low-temperature orthorhombic structure belongs to the space group Fmmm (point group $D_{2h}$). Freshly cleaved CaFe$_2$As$_2$ and EuFe$_2$As$_2$ single crystals have been measured in a quasi-back-scattering geometry from the $ab$ surface. The 476-and 647-nm laser beams of a Kr$^+$ ion laser with a total incident power smaller than 15~mW were focused to a 50$\times$125~$\mu$m$^2$ spot on the sample surface. The scattered light was analyzed by a triple grating spectrometer and collected by a liquid N$_2$-cooled CCD detector. The data were corrected for the spectral response of the system at different wavelengths~\cite{Blumberg_PRB1994}. The polarization configuration ($e^Ie^S$) is defined by the polarization of the incident and scattered photons, $e^I$ and $e^S$, respectively. The polarization vectors $x$ = $[110]$ and $y$ = $ [\bar110]$ are along the nearest Fe-Fe bonds corresponding, respectively, to the antiferromagnetic (AFM) and ferromagnetic (FM) directions in the SDW phase, and $X$ = $[100]$ and $Y$ = $[010]$ form a $45^\circ$ angle with them.


\begin{figure}[!t]
\begin{center}
\includegraphics[width=1.0\columnwidth]{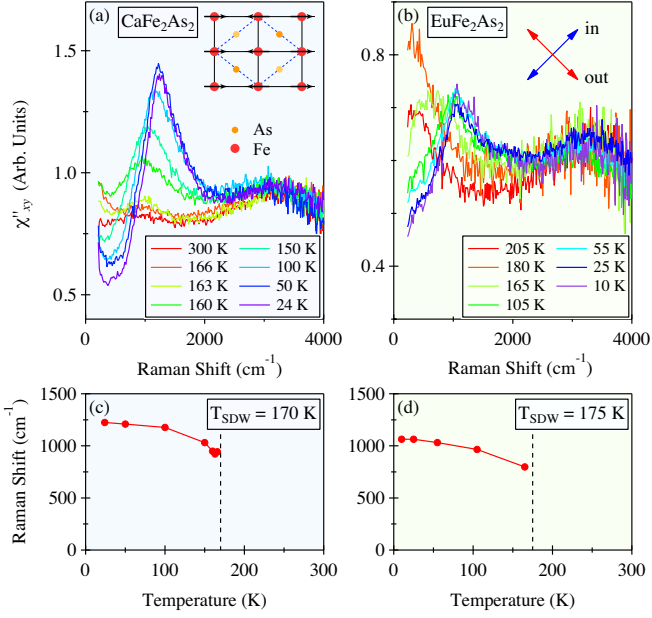}
\end{center}
 \caption{\label{Fig: temperature} (Color online) (a) Temperature dependence of the Raman response of untwinned CaFe$_2$As$_2$ in the $XY$ polarization configuration corresponding to the $B_{2g}$($D_{4h}$)/$A_g$($D_{2h}$) symmetry channel, recorded with a 647-nm laser excitation. The inset illustrates one Fe-As layer of the four-Fe unit cell in the low-temperature phase, along with the Fe magnetic moments (red arrow). (b) Same as (a) but for twinned EuFe$_2$As$_2$. The inset shows the incident and scattered light polarizations with respect to the lattice orientation in the inset of (a). (c) and (d) Frequency of the coherence peak as a function of temperature for CaFe$_2$As$_2$ and EuFe$_2$As$_2$, respectively.}
\end{figure}

\begin{figure}[!t]
\begin{center}
 \includegraphics[width=1.0\columnwidth]{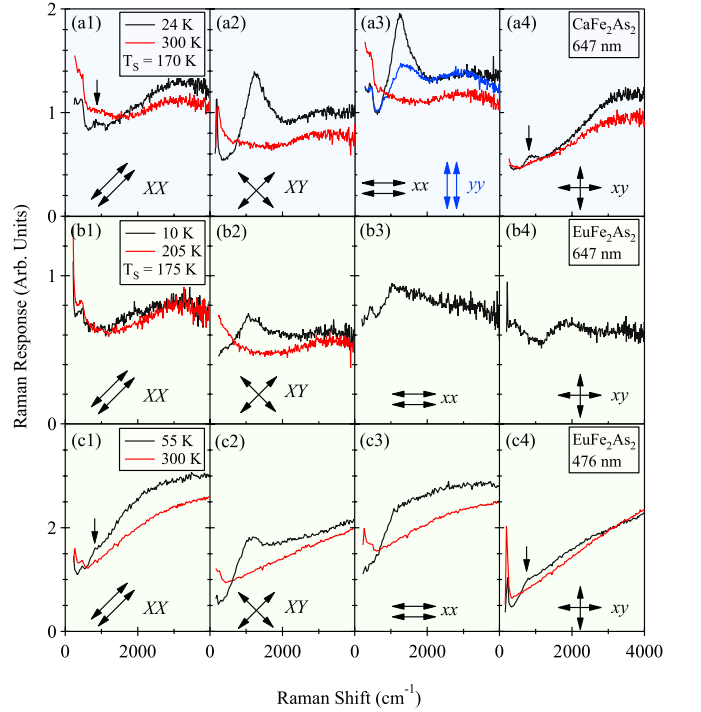}
\end{center}
 \caption{\label{Fig: data} (Color online) (a1)-(a4) Raman response of CaFe$_2$As$_2$ measured with 647-nm laser excitation in different in-plane polarization configurations (denoted by the arrows) in the paramagnetic state (300~K, red) and in the detwinned SDW phase (24~K, black/blue). (b1)-(b4) Raman response of EuFe$_2$As$_2$ in different in-plane polarization configurations in the paramagnetic state (205~K, red) and the twinned SDW state (10~K, black). (c1)-(c4) Same as (b1)-(b4) with the 476-nm laser excitation.}
\end{figure}

In Fig.~\ref{Fig: temperature}, we show the Raman response of CaFe$_2$As$_2$ and EuFe$_2$As$_2$ for the $XY$ scattering geometry at various temperatures above and below $T_{SDW}$. 
We observe a spectral weight transfer from low frequency to above 800~cm$^{-1}$ with the development of a coherence peak. As indicated in Figs. \ref{Fig: temperature}(c) and \ref{Fig: temperature}(d), the coherence peak in CaFe$_2$As$_2$ (EuFe$_2$As$_2$) hardens from 920 (800) to 1220 (1060)~cm$^{-1}$ from $T_{SDW}$ to the lowest measured temperature. This spectral feature was previously assigned to the formation of a SDW gap \cite{Chauviere_PRB2011}, which is accompanied by the appearance of a Dirac cone in the electronic structure \cite{Richard_PRL2010,Harrison_PRB2009}. 

\begin{figure*}[!t]
\begin{center}
 \includegraphics[width=2\columnwidth]{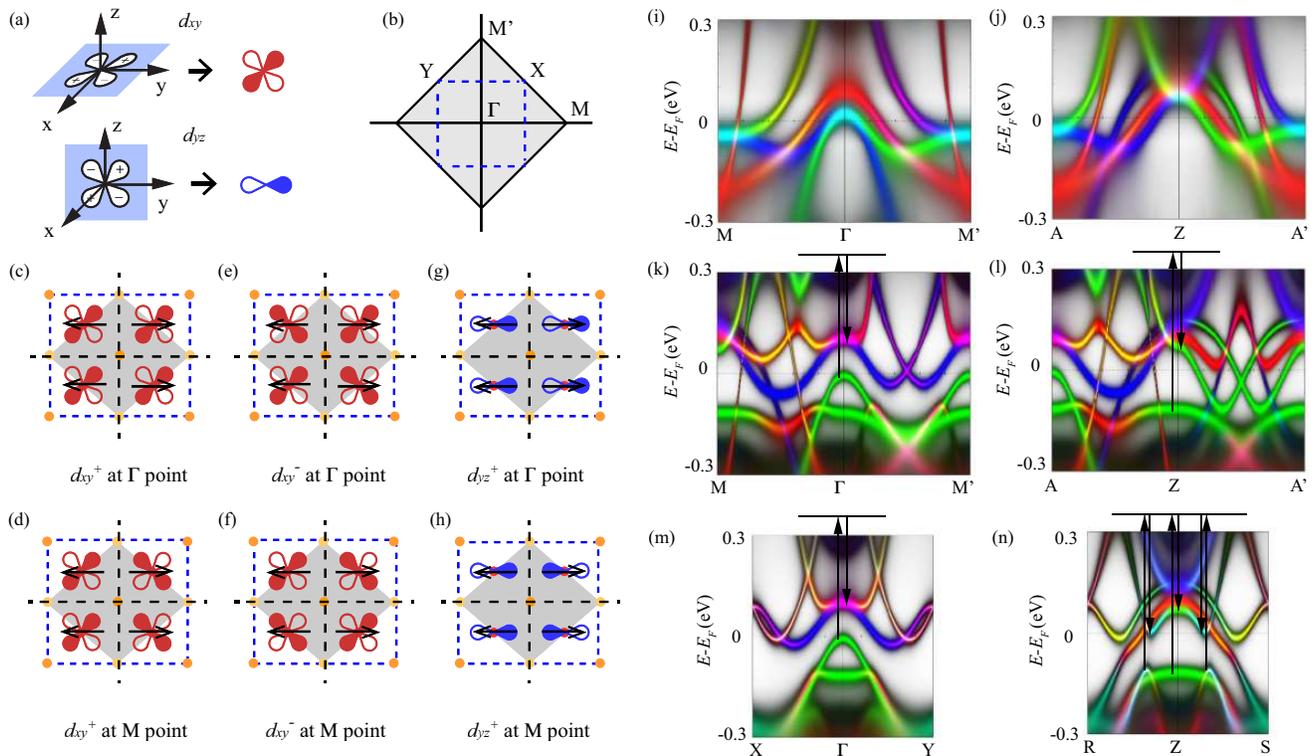}
\end{center}
 \caption{\label{Fig: discussion}(Color online) (a) Diagram of the $d_{xy}$ and $d_{yz}$ orbitals. (b) High-temperature 2-Fe BZ (gray shade) and low-temperature folded BZ (dashed line). (c-h) Schematic representations of Fe 3$d_{xy}^+$, $d_{xy}^-$, and $d_{yz}^+$ orbitals in real space at the high-symmetry points $\Gamma$ and $M$ in the momentum space. Filled and empty lobes represent the sign of the wave functions. The dashed blue lines show the four-Fe unit cell in the SDW phase. The dashed black lines are the $xz$ or $yz$ mirror planes. The black arrows represent the local moment in the SDW phase. A complete schematic of wave function phase change of the three dominant orbitals in high-symmetry points $\Gamma$, $M$ and $M$$^\prime$ is shown in the Appendix. (i)-(n) Single-particle spectral function of CaFe$_2$As$_2$ in the paramagnetic phase ((i) and (j)) and SDW phase ((k)-(n)) along high-symmetry lines in the $k_z$ = 0 (left column) and $k_z$ = 2$\pi$/$c$ (right column row) planes. The $d_{xy}$, $d_{xz}$, and $d_{yz}$ orbital characters are denoted in red, green, and blue, respectively. The black arrows in (k)-(n) mark the main contributing transitions.}
\end{figure*}

We performed polarization-dependent measurements on CaFe$_2$As$_2$ and EuFe$_2$As$_2$ samples. The electronic states in the vicinity of the Fermi level ($E_F$) in the iron-pnictides consists mainly of Fe 3$d$ orbitals. Locally, these states decompose into $A_{1g}$ ($d_{z^2}$), $B_{2g}$ ($d_{x^2-y^2}$), $B_{1g}$ ($d_{xy}$), and two $E_{g}$ ($d_{xz}$, $d_{yz}$) states, which behave differently under the symmetry operations of the $D_{4h}$ group. 
We note that in the low-temperature SDW phase the crystal structure of $A$Fe$_2$As$_2$ is lowered to the $D_{2h}$ group and the irreducible representations change. For example, $A_{1g}$($D_{4h}$) and $B_{2g}$($D_{4h}$) merge into $A_{g}$($D_{2h}$), and $A_{2g}$($D_{4h}$) and $B_{1g}$($D_{4h}$) merge into $B_{1g}$($D_{2h}$). We list all in-plane polarization configurations and the symmetry channels they can couple to in both the paramagnetic and SDW phases in Table~\ref{Table: merge}, from which the evolution of the $D_{4h}$ and $D_{2h}$ irreducible representations are directly shown.

\begin{table}[h]
\renewcommand\arraystretch{1.2}
\caption{Light polarization configurations and corresponding symmetry channels probed  in the paramagnetic and SDW phases.}
\label{Table: merge}
\centering
\begin{tabular}{ccc}
\hline\hline
 \diagbox{Pol.}{phase}&\,\,\,\,Paramagnetic & SDW\\
 \hline
\multirow{2}{*}{\textit{XX}}& \multirow{2}{*}{$A_{1g}$+$B_{1g}$} & \,\,$A_g$+$B_{1g}$\\
&&Not a proper geometry\\
\multirow{2}{*}{\textit{XY}}& \multirow{2}{*}{$A_{2g}$+$B_{2g}$} & \,\,$A_g$+$B_{1g}$\\
&&Not a proper geometry\\
\textit{xx, yy} &$A_{1g}$+$B_{2g}$&$A_g$\\
\textit{xy} &$A_{2g}$+$B_{1g}$&$B_{1g}$\\
\hline
\hline
\end{tabular}

\label{default}
\end{table}%

Our results are displayed in Fig. \ref{Fig: data}. We first discuss the Raman responses obtained on CaFe$_2$As$_2$ (first row in Fig. \ref{Fig: data}). At room temperature, we observe that a continuum extends to the highest measured frequency in all polarization configurations. Typically, a twinned orthorhombic structure forms upon cooling down the samples below $T_{SDW}$, as can be viewed under a microscope with crossed-polarized light~\cite{Prozorov_PRB2009}. To maximize the size of the domains, our samples were cooled down in two steps. At temperatures way above $T_{SDW}$, the samples were cooled at a rate higher than 60~K/hr. Upon approaching $T_{SDW}$ though, this cooling rate was decreased below 1~K/hr. Using this procedure, a mono-domain in millimeter size formed in our CaFe$_2$As$_2$ samples, which we confirmed using laser illumination with $XY$ polarization. In Figs.~\ref{Fig: data}(a1)-(a4) we show the Raman scattering intensity of a CaFe$_2$As$_2$ sample slowly cooled down. The coherence peak around 1220~cm$^{-1}$ only appears in the $XY$, $xx$, and $yy$ polarization spectra at low temperature. 
Moreover, a large anisotropy of intensity for this coherence peak is observed between the $xx$ and $yy$ polarizations, even though they share the same low-frequency phonon spectra and high-frequency response.
Indeed, the intensity of the peak in the $yy$ polarization is less than half of that in the $xx$ polarization, and the frequency is slightly higher. 
In addition, for the $XX$ and $xy$ polarizations, we distinguish clearly a smaller feature at 830~cm$^{-1}$ that has not been reported in previous Raman studies.  

We now switch to our results on EuFe$_2$As$_2$. Although we did not succeed in detwinning completely the sample using the same cooling procedure, similar physics is observed. The paramagnetic state spectra in the $XX$ and $XY$ configurations are very similar to those obtained on CaFe$_2$As$_2$. As with CaFe$_2$As$_2$, a strong coherence peak is also detected in EuFe$_2$As$_2$ under the $XY$ and $xx$ (or $yy$) configurations, albeit for a slightly smaller peak frequency (1060~cm$^{-1}$). In order to confirm the Raman nature of the features observed, we measured Raman spectra on the same sample using a 476-nm laser excitation. The results, displayed in Figs.~\ref{Fig: data}(c1)-\ref{Fig: data}(c4), show different backgrounds as compared to the ones recorded with 647-nm light. In particular, the broad peaks found at 1800 and 3200~cm$^{-1}$ in the 647-nm spectra are absent in the 476-nm spectra, suggesting that they possibly correspond to luminescence signals. 
Interestingly, the smaller feature found in CaFe$_2$As$_2$ using the $XX$ and $xy$ polarizations is only present under 476-nm excitation.

In order to understand the origin of these electronic Raman excitations, we calculated the band dispersion of CaFe$_2$As$_2$ by using DFT+DMFT in the paramagnetic and SDW phases~\cite{Yin_naturephysics2011}. We used the experimental lattice constants and internal coordinates of the paramagnetic phase. The Coulomb interaction and the Hund's coupling are from Ref.~\cite{Yin_Natmat2011}. The band dispersions [single-particle spectral function $A(k,\omega)]$ along high-symmetry lines are shown in Figs.~\ref{Fig: discussion}(i)-(n) with the orbital characters represented by different colors. The high-symmetry points are labeled in Fig.~\ref{Fig: discussion}(b).

The presence of different gap energies in two different symmetry channels in the SDW phase arises from the complexity of the folded band structures. 
The band structure near $E_F$ is dominated by the $d_{xy}$, $d_{xz}$ and $d_{yz}$ orbitals of Fe. 
In the high-temperature phase, there are three hole FSs around the $\Gamma$ point and two electron FSs around the M point. 
There is a SDW order that not only causes the lifting of the degenerated $d_{xz/yz}$ bands but also induces Brillouin-zone (BZ) folding from $A(\pi/a, 0, 2\pi/c)$ to $\Gamma$. At the $M$ ($M^\prime)$ point, the two electron bands hybridize with the three folded hole bands from $Z~(\Gamma$) and a band gap is opened around $M~(Z)$. In contrast to the Dirac node at $E_F$ reported in BaFe$_2$As$_2$ \cite{Richard_PRL2010,Harrison_PRB2009}, the Dirac point in CaFe$_2$As$_2$ is 45~meV above $E_F$. Interestingly, there is a second Dirac point at $E_F$ along the FM direction in the $k_z=0$ plane between the $d_{yz}$ band and the folded $d_{yz}$ band. In the $k_z=2\pi/c$ plane, the second Dirac cone carries the $d_{xz}$ characters and its energy is 36~meV below $E_F$.

Raman scattering is a small $q$ process since the momentum conservation requires the momentum transfer between the initial and final states to be equal to the momentum difference of the scattered and incident lights. 
As with infrared absorption, the interband electronic Raman scattering intensity is determined by the joint density of states and the coherence factor. In the case of non-resonance scattering, the scattering intensity is a reflection of the density of states of the initial and final states.

\begin{figure}[!t]
\begin{center}
\includegraphics[width=1\columnwidth]{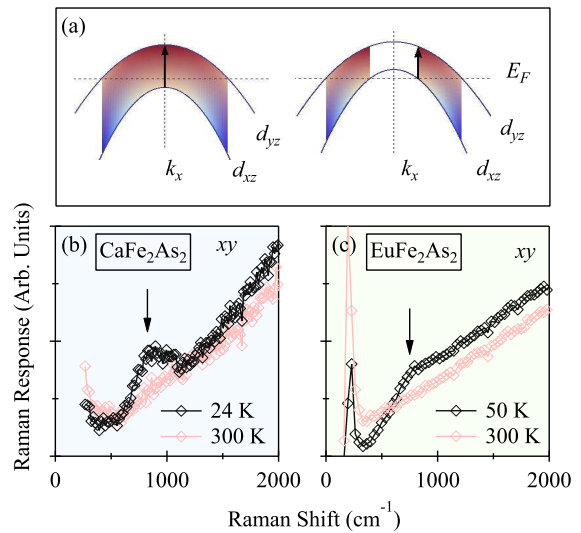}
\end{center}
\caption{\label{Fig: B1g}(Color online) (a) $d_{xz}$-$d_{yz}$ transition at the BZ center and $k_F$. The excitations are allowed in the shaded areas. (b) and (c) Differential Raman spectra between 300 and 24 K (50 K) in CaFe$_2$As$_2$ and EuFe$_2$As$_2$, respectively. The peak in (b) and (c), denoted by arrows, correspond to the excitations at the momentum indicated by arrows in (a).}
\end{figure}


The symmetries of the Fe orbitals are derived from the symmetries of their wave functions. Away from the $\Gamma$ point, the phase term $e^{ik\cdot R}$ in the wave functions can change the parity of some bands~\cite{Brouet_PRB2012}. For example, we illustrate in Figs.~\ref{Fig: discussion}(c)-(h) the $d_{xy}^+$,  $d_{xy}^-$, and $d_{yz}^+$ orbitals at the $\Gamma$ and $M$ point. At the $M$ point, the parity to the $yz$ mirror plan is the same as at the $\Gamma$ point, and the parity to the $xz$ mirror plane becomes the opposite. 
We summarize the symmetry of the three dominant orbitals in the paramagnetic and SDW phases in Table~\ref{Table: merge}. A complete symmetry analysis is shown in the Appendix.

\begin{table}[h]
\renewcommand\arraystretch{1.2}
\caption{Symmetries of $d_{xy}$, $d_{yz}$ and $d_{xz}$ orbitals at high-symmetry momentum space points, and the allowed transitions in each symmetry channel in the paramagnetic and SDW phases.}
\label{Table: symmetry}
\centering
\begin{tabular}{{p{35mm}}*{6}{p{7mm}}}
\hline\hline
\multicolumn{7}{c}{High-temperature paramagnetic phase, $D_{4h}$}\\

Momentum/orbitals& $d_{xy}^+$ & $d_{xy}^-$&$d_{yz}^+$&$d_{yz}^-$&$d_{xz}^+$& $d_{xz}^-$\\
 \hline
$\Gamma$ & $B_{2g}$ &$A_{2u}$&$E_{g}$& $E_{u}$ &$E_{g}$&$E_{u}$\\

$M$ &$E_{u}$ &$E_{g}$&$A_{2u}$& $A_{2g}$ &$A_{1u}$&$A_{1g}$\\

$M^\prime$&$E_{g}$ &$E_{u}$&$A_{2g}$& $A_{2u}$ &$A_{1g}$&$A_{1u}$\\
Symmetry channel&\multicolumn{6}{c}{allowed transitions}\\
\hline
$A_{1g}$&\multicolumn{6}{c}{intra-orbital}\\
$A_{2g}$&\multicolumn{6}{c}{$d_{xz}^+$~$\leftrightarrow$~$d_{yz}^+$, $d_{xz}^-$~$\leftrightarrow$~$d_{yz}^-$}\\
$B_{1g}$&\multicolumn{6}{c}{-}\\
$B_{2g}$&\multicolumn{6}{c}{-}\\

\multicolumn{7}{c}{Low-temperature SDW phase, $D_{2h}$}\\

Momentum/orbitals & $d_{xy}^+$ & $d_{xy}^-$&$d_{yz}^+$&$d_{yz}^-$&$d_{xz}^+$& $d_{xz}^-$\\
 \hline
$\Gamma$ & $B_{1g}$ &$B_{1u}$&$B_{3g}$& $B_{3u}$ &$B_{2g}$&$B_{2u}$\\

$M$ &$B_{3u}$ &$B_{3g}$&$B_{1u}$& $B_{1g}$ &$A_{u}$&$A_{g}$\\

$M^\prime$&$B_{3g}$ &$B_{3u}$&$B_{1g}$& $B_{1u}$ &$A_{g}$&$A_{u}$\\

Symmetry channel&\multicolumn{6}{c}{allowed transitions}\\
\hline
$A_g$&\multicolumn{6}{c}{intra-orbital}\\
$B_{1g}$&\multicolumn{6}{c}{$d_{xz}^+$~$\leftrightarrow$~$d_{yz}^+$, $d_{xz}^-$~$\leftrightarrow$~$d_{yz}^-$}\\
\hline
\hline
\end{tabular}
\end{table}

The selection rules between these orbitals are determined by the product of the initial-and final-state symmetries. For the high-temperature paramagnetic phase, only the intra-orbital transitions are allowed in the $A_{1g}$ symmetry channel, while for the SDW phase, besides the intra-orbital transitions in the $A_{1g}$ symmetry channel,  the $d_{xz}^+$ $\leftrightarrow$ $d_{yz}^+$ and $d_{xz}^-$ $\leftrightarrow$ $d_{yz}^-$ transitions become Raman active in the $B_{1g}$ symmetry channel. The selection rules for each in-plane symmetry channel in both the paramagnetic and SDW phases are listed in Table~\ref{Table: merge}. 
The Raman scattering vertices in the $B_{1g}$ and $B_{2g}$ symmetries for the one-Fe BZ for each pair of the orbital characters of the initial and final states are calculated in~\cite{Brouet_PRB2013}.

While phononic Raman scattering requires as a general selection rule that the symmetry of the phonon states be chosen by the incident and scattered light polarizations, 
electronic Raman scattering requires more complex selection rules for the symmetry of the initial and final states. 
The light polarized along the AFM direction ($x$) can excite $d_{xz}$ bands. For light polarized along the FM ($y$) direction, only $d_{yz}$ can be excited. 
Therefore, the peak in the $xx$ polarization can only arise from $d_{xz}$~$\leftrightarrow$~$d_{xz}$ transitions.
Our calculations further confirm that the $A_g$ symmetry excitation primarily comes from the $Z$ point, where a band gap opens between the original and the folded $d_{xz}$ bands (Fig.~\ref{Fig: discussion}(l)). The second Dirac point in the $Z-R$ and $Z-S$ directions also has a contribution to the Raman spectra~(Fig.~\ref{Fig: discussion}(n)). Since the second Dirac cone contains $d_{yz}$ components, the scattering for $yy$ polarization is not fully prohibited. The main contributing transitions are marked with the black arrows in Figs.~\ref{Fig: discussion}

The small feature in the $B_{1g}$ channel arises from the $d_{xz}$~$\leftrightarrow$~$d_{yz}$ transition at the $\Gamma$ point. We illustrate in Fig.~\ref{Fig: B1g}(a) this transition for two different levels of $E_F$. When the $d_{xz}$ band is totally below $E_F$, the transition will undergo at the BZ center, where the density of states is maximum, resulting in a strong scattering signal. This situation is met at the $\Gamma$ point for both materials we measured (Fig.~\ref{Fig: B1g}(b) and (c))~\cite{Kondo_PRB2010,Richard_JPCM2014}. 
As a comparison, when the $d_{xz}$ band crosses $E_F$, the minimum energy transfer is at the momentum wave vector $k_F$.  In this situation, the spectrum shows a threshold at this minimum energy.

Interestingly, the Raman data are consistent with the $d_{xz}/d_{yz}$ splitting estimated from angle-resolved photoemission spectroscopy (ARPES) data~\cite{YiM_PNAS2011,Richard_JPCM2014}, thus reinforcing its interpretation. Unlike ARPES though, which can access the unoccupied states only for a very small energy range, polarized electronic Raman scattering allows us to probe directly the $d_{xz}/d_{yz}$ splitting at the $\Gamma$ point, across $E_F$.

In conclusion, we reported temperature-dependent and polarization-resolved Raman scattering on twinned CaFe$_2$As$_2$ and EuFe$_2$As$_2$ and detwinned mono-domain CaFe$_2$As$_2$ single crystals. Two spectral features are observed in two different symmetry channels in the SDW phase. Based on symmetry arguments and DFT+DMFT calculations of the orbital-resolved electronic band structure, we identified these transitions. In the $A_g$ symmetry channel, there is a spectral weight transfer from low-frequency and the formation of a coherence peak around 1200 cm$^{-1}$, which arises from the SDW band-folding-induced intra-orbital transition at the $Z$ point and near the second Dirac point. Moreover, the coherence peak is anisotropic for light polarizations along the AFM and FM directions in detwinned CaFe$_2$As$_2$, directly revealing the inequivalent occupancy of the $d_{xz}$ and $d_{yz}$ orbitals. 
In the $B_{1g}$ symmetry channel, a Raman scattering peak around 800 cm$^{-1}$ arises from a transition between $d_{xz}$ and $d_{yz}$ at the $\Gamma$ point and reveals the lifted degeneracy of these two orbitals.

W.-L.Z. acknowledges ICAM (NSF-IMI Grant No. DMR-0844115) and NSF (Grant No. DMR-1104884).
P.R. and H.D. acknowledge MoST (Grants No. 2011CBA001001 and No. 2015CB921301) and National Natural Science Foundation of China (Grant No. 11274362) of China. Z.P.Y. acknowledges NSF Grant No. DMR130814. Z.B. acknowledges the NCN, Poland (Grant No. 2011/01/B/ST5/06937). A.S.S. and G.B. acknowledge the U.S. Department of Energy, BES, and Division of Materials Sciences and Engineering under Awards to ORNL and Grant No. DE-SC0005463 correspondingly.

\appendix*{
\section{Symmetry analysis}

In this section we show the details of the orbital symmetry analysis. In Fig. A1, we plot the six dominant Fe $3d$ orbitals in two different Fe-As layers in a unit cell from the top view at the $\Gamma$, $M$ and $M^\prime$ point. Away from the $\Gamma$ point, the wave function has a phase change of $e^{ik\cdot R}$. The parities of these orbitals under all $D_{4h}$ and $D_{2h}$ symmetry operations are listed in Table \ref{Table: symmetry} and \ref{Table: symmetry2}.

\begin{figure*}[!t]
\begin{center}
\includegraphics[width=2\columnwidth]{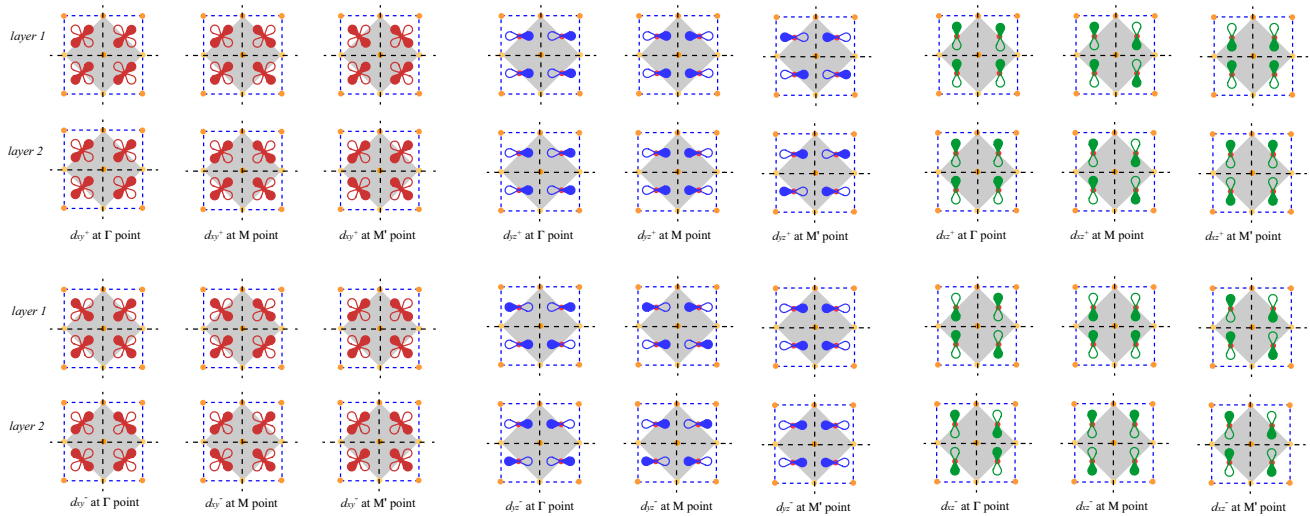}
\end{center}
\caption{\label{Fig: A1} (Color online) $d_{xy}$, $d_{yz}$, and $d_{yz}$ orbitals in two different Fe-As layers in a unit cell in real space at high-symmetry points $\Gamma$, $M$ and $M^\prime$ in the momentum space.  }
\end{figure*}

\begin{table*}[h]
\renewcommand\arraystretch{1.2}
\caption{Symmetries of $d_{xy}$, $d_{yz}$, and $d_{xz}$ orbitals at high-symmetry momentum space points in the SDW phase.}
\label{Table: symmetry}
\centering
\begin{tabular}{*{13}{c}}
\hline\hline
\multicolumn{13}{c}{High-temperature paramagnetic phase, $D_{4h}$}\\
\hline
{Orbitals}&Momentum& $E$ & $C_4(z)$&$C_2$&$C_2^\prime$&$C_2^{\prime\prime}$&$i$&$S_4$&$\sigma_h$&$\sigma_v$&$\sigma_d$&symmetry\\

$d_{xy}^+$&$\Gamma$&1&-1&1&-1&1&1&-1&1&-1&1&$B_{2g}$\\
$d_{xy}^+$&$M$&2&0&-2&0&0&-2&0&2&0&0&$E_{u}$\\
$d_{xy}^+$&$M^\prime$&2&0&-2&0&0&2&0&-2&0&0&$E_{g}$\\

$d_{xy}^-$&$\Gamma$ & 1 &1&1& -1&-1&-1&-1&-1&1&1&$A_{2u}$\\
$d_{xy}^-$&$M$ &2&0&-2&0&0&2&0&-2&0&0&$E_{g}$\\
$d_{xy}^-$&$M^\prime$&2&0&-2& 0&0&-2&0&2&0&0&$E_{u}$\\

$d_{yz}^+$&$\Gamma$ &2&0&-2&0&0&2&0&-2&0&0&$E_{g}$\\
$d_{yz}^+$&$M$ &1&1&1&-1&-1&-1&-1&-1&1&1&$A_{2u}$\\
$d_{yz}^+$&$M^\prime$ &1&1&1&-1&-1&1&1&1&-1&-1&$A_{2g}$\\

$d_{yz}^-$&$\Gamma$ &2&0&-2& 0&0&-2&0&2&0&0&$E_{u}$\\
$d_{yz}^-$&$M$ &1&1&1&-1&-1&1&1&1&-1&-1&$A_{2g}$\\
$d_{yz}^-$&$M^\prime$&1&1&1&-1&-1&-1&-1&-1&1&1&$A_{2u}$\\

$d_{xz}^+$&$\Gamma$&2&0&-2&0&0&2&0&-2&0&0&$E_{g}$\\
$d_{xz}^+$&$M$ &1&1&1&1&1&-1&-1&-1&-1&-1&$A_{1u}$\\
$d_{xz}^+$&$M^\prime$ & 1 &1&1& 1&1&1&1&1&1&1&$A_{1g}$\\

$d_{xz}^-$&$\Gamma$  &2&0&-2& 0&0&-2&0&2&0&0&$E_{u}$\\
$d_{xz}^-$&$M$  & 1 &1&1& 1&1&1&1&1&1&1&$A_{1g}$\\
$d_{xz}^-$&$M^\prime$ &1&1&1&1&1&-1&-1&-1&-1&-1&$A_{1u}$\\

\hline
\hline
\end{tabular}
\end{table*}

\begin{table*}[h]
\renewcommand\arraystretch{1.2}
\caption{Symmetries of $d_{xy}$, $d_{yz}$, and $d_{xz}$ orbitals at high-symmetry momentum space points in the SDW phase.}
\label{Table: symmetry2}
\centering
\begin{tabular}{*{11}{c}}
\hline\hline
\multicolumn{11}{c}{Low-temperature SDW phase, $D_{2h}$}\\

{Orbitals}&Momentum& $E$ & $C_2(z)$&$C_2(y)$&$C_2(x)$&$i$&$\sigma_{xy}$&$\sigma_{xz}$&$\sigma_{yz}$&symmetry\\
\hline
$d_{xy}^+$&$\Gamma$&1&1&-1&-1&1&1&-1&-1&$B_{1g}$\\
$d_{xy}^+$&$M$&1&-1&-1&1&-1&1&1&-1&$B_{3u}$\\
$d_{xy}^+$&$M^\prime$&1&-1&-1&1&1&-1&-1&1&$B_{3g}$\\

$d_{xy}^-$&$\Gamma$&1&1&-1&-1&-1&-1&1&1&$B_{1u}$\\
$d_{xy}^-$&$M$&1&-1&-1&1&1&-1&-1&1&$B_{3g}$\\
$d_{xy}^-$&$M^\prime$&1&-1&-1&1&-1&1&1&-1&$B_{3u}$\\
$d_{yz}^+$&$\Gamma$&1&-1&-1&1&1&-1&-1&1&$B_{3g}$\\
$d_{yz}^+$&$M$&1&1&-1&-1&-1&-1&1&1&$B_{1u}$\\
$d_{yz}^+$&$M^\prime$&1&1&-1&-1&1&1&-1&-1&$B_{1g}$\\
$d_{yz}^-$&$\Gamma$&1&-1&-1&1&-1&1&1&-1&$B_{3u}$\\
$d_{yz}^-$&$M$&1&1&-1&-1&1&1&-1&-1&$B_{1g}$\\
$d_{yz}^-$&$M^\prime$&1&1&-1&-1&-1&-1&1&1&$B_{1u}$\\
$d_{xz}^+$&$\Gamma$&1&-1&1&-1&1&-1&1&-1&$B_{2g}$\\
$d_{xz}^+$&$M$&-1&-1&-1&-1&-1&-1&-1&-1&$A_{u}$\\
$d_{xz}^+$&$M^\prime$&1&1&1&1&1&1&1&1&$A_{g}$\\
$d_{xz}^-$&$\Gamma$&1&-1&1&-1&-1&1&-1&1&$B_{2u}$\\
$d_{xz}^-$&$M$&1&1&1&1&1&1&1&1&$A_{g}$\\
$d_{xz}^-$&$M^\prime$&-1&-1&-1&-1&-1&-1&-1&-1&$A_{u}$\\

\hline
\hline
\end{tabular}
\end{table*}
}


%

\end{document}